\RequirePackage[final]{graphicx}
\documentclass[aps, superscriptaddress,pra,twocolumn,footinbib]{revtex4-1}
\usepackage[utf8]{inputenc}
\usepackage{amsmath}
\usepackage{graphicx}
\usepackage{fixme}
\usepackage{color}

\begin{document}

\title{Topological superfluidity of lattice  fermions  inside  a Bose-Einstein condensate}
\author{Jonatan Melk\ae r Midtgaard}
\affiliation{Department of Physics and Astronomy,  Aarhus University, Ny Munkegade, DK-8000 Aarhus C, Denmark}
\author{Zhigang Wu}
\affiliation{Department of Physics and Astronomy,  Aarhus University, Ny Munkegade, DK-8000 Aarhus C, Denmark}
\author{G.\ M.\ Bruun}
\affiliation{Department of Physics and Astronomy,  Aarhus University, Ny Munkegade, DK-8000 Aarhus C, Denmark}
\date{\today}

\begin{abstract}
We calculate the phase diagram of identical fermions in a 2-dimensional (2D) lattice immersed in a 3D Bose-Einstein condensate (BEC). The fermions exchange density fluctuations in the 
BEC, which gives rise to  an attractive induced interaction. The resulting zero temperature phase diagram exhibits topological $p_x+ip_y$ superfluid phases as well as a phase separation 
region. We show how to use the flexibility of the Bose-Fermi mixture  to tune the induced interaction, so that it maximises the pairing between nearest neighbour sites, whereas phase 
separation originating from long range interactions is suppressed.  Finally, we calculate the Berezinskii-Kosterlitz-Thouless (BKT) critical temperature of the topological superfluid in the lattice
 and    discuss  experimental realisations.  
\end{abstract}
\pacs{$\ldots$}

\maketitle

\section{Introduction}
Ever since topological superfluids/superconductors with Majorana modes were  predicted to exist~\cite{Kitaev2001}, it has been a major research goal to 
 detect them. Experimental evidence for Majorana modes have been 
reported in 1D wires~\cite{Mourik2012,Deng2012,Das2012,Rokhinson2012,Finck2013,NadjPerge2014,Albrecht2016}, 
and  Sr$_2$RuO$_4$ is a promising candidate for realising a 2D topological superconductor~\cite{Ishida1998,Nelson2004,Kidwingira2006,Xia2006}. 
An unambiguous detection of topological superconductivity/superfluidity is however still lacking, partly due to the complexity 
of these condensed matter systems. There are several proposals to use the clean and highly flexible atomic quantum gases to  overcome this
difficulty. The first scheme was based on Fermi gases interacting via a $p$-wave Feshbach resonance~\cite{Gurarie2007}, but these 
systems  suffer from short lifetimes~\cite{Gunter2005,Gaebler2007,Levinsen2007,JonaLasinio2008}. 
Other quantum gas proposals include schemes based on optical lattices~\cite{Buhler2014,Massignan2010,Mathey2006,Wu2012}, 
synthetic spin-orbit coupling~\cite{Zhang2008,Sato2009,Jiang2011}, driven dissipation~\cite{Bardyn2012,Diehl2011}, dipolar molecules~\cite{Cooper2009,Liu2012,Fedorov2016}, and 
mixed dimension Fermi-Fermi mixtures~\cite{Nishida2009}. Unfortunately,  none of these systems  have  been realised so far. 

Very recently, two of us (ZW and GMB) demonstrated that a mixed dimension Fermi-Bose mixture constitutes a  promising system to realise a 2D topological superfluid~\cite{Wu2016}. 
 In this proposal, identical fermions are confined in a 2D plane and they interact via density modulations in a surrounding BEC. Due to the 
 high compressibility of the BEC, this induced interaction is  strong, and one can moreover control the range by varying the BEC coherence 
 length. This flexibility can be used to make the critical temperature of the topological superfluid  high, 
 while keeping three-body losses small. The purpose of the present paper is to examine this promising scheme in a setup, where the fermions are moving in a 2D optical 
 lattice. Optical lattices offer the particular advantage of single site resolution spectroscopy~\cite{Sherson2010,Bakr2009}, which presents unique opportunities to 
 detect and manipulate Majarona edge modes~\cite{Nascimbene2013,Goldman2016}. We therefore investigate the phase diagram of identical
  fermions moving in a 2D square lattice immersed in a 3D BEC, which gives rise to 
  an attractive induced interaction between the  fermions. As a result, the ground state of the fermions is a topological $p_x+ip_y$ superfluid, but the system also 
  suffers from a phase separation instability  originating from an underlying particle-hole symmetry in  the lattice. Taking the competition of these two instabilities into account, we 
  calculate the zero temperature phase diagram of the fermions as a function of the Bose-Fermi coupling strength and the  filling fraction.
   We show how to use the flexibility of the Bose-Fermi system to 
  tune the interaction in order to maximise the pairing strength, while keeping the system stable against phase separation. The key point to solve this delicate problem turns out to be to 
  adjust the range of the interaction by changing the BEC coherence length, so that it induces pairing on neighboring sites, while it does not lead to a significant long range interaction effects beyond 
  nearest neighbors. We finally calculate the BKT critical temperature for the superfluid phase and    discuss  experimental realisations.

\section{Model}
\begin{figure}[htb]
\centering
\includegraphics[width=\columnwidth]{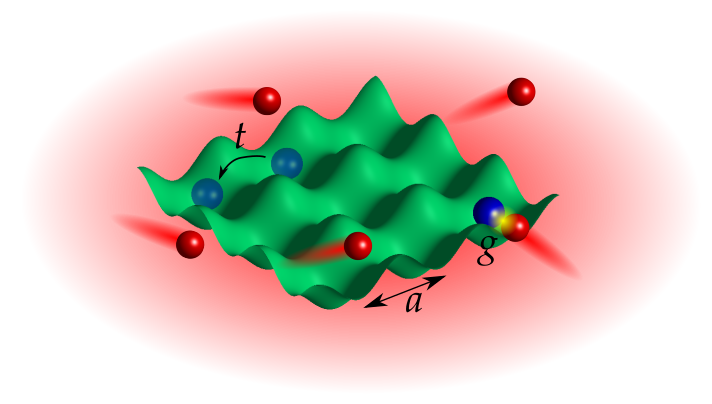}
\caption{(Color online). Illustration of the mixed-dimensional system. The fermions (blue balls) move in a 2D square lattice in the $xy$-plane with lattice constant $a$ and hopping matrix element $t$.
The lattice is immersed in a 3D BEC (red balls) and the fermions and bosons interact through a contact interaction with strength $g$.}
\label{fig:system}
\end{figure}

We consider fermionic atoms of mass $m$ moving in a 2D square  lattice  in the $xy$-plane. The lattice is immersed 
in a 3D BEC consisting of atoms with mass $m_\text{B}$, which is weakly  interacting with $n_0a^3_\text{B}\ll 1$, so that it can be accurately described by Bogoliubov theory. Here $n_0$ denotes
 the condensate density and $a_\text{B}$ the boson-boson scattering length. The setup is  illustrated in  Fig. \ref{fig:system}. The 
%%%%%%%%%%%%%%%%%%%%%%%%%%%%%%%%%%%%%%%%%%%%%%%%%%%%%%%%%%%%%%
%%%%%%%%%%%%%%%%%%%%%%%%%%%%%%%%%%%%%%%%%%%%%%%%%%%%%%%%%%%%%%
 Hamiltonian is
\begin{align}
H=& -t \sum_{\langle i,j \rangle}  (a_i^\dagger a_j+\text{h.c.})
+\sum_{\mathbf k}E_{\mathbf k} \gamma^\dagger_{\mathbf k} \gamma_{\mathbf k}+H_\text{int},
\label{Hint}
\end{align}
where $t$ is the hopping matrix element of the fermions between nearest neighbor sites $\langle i,j \rangle$, $a_i^\dagger$ creates a fermion at lattice site $i$, and 
$\gamma^\dagger_{\mathbf k}$ creates a Bogoliubov mode in the BEC with momentum $\mathbf k$ and energy 
$E_{\mathbf k}=[\epsilon_k^\text{B}(\epsilon_k^\text{B}+8\pi n_0a_\text{B}/m_\text{B})]^{1/2}$ where $\epsilon_k^\text{B}=k^2/2m_\text{B}$. We use units for 
which $k_B=\hbar=1$. Within the 
pseudopotential approximation, the interaction between the fermions and the bosons is 
\begin{align}
H_{\text{int}} = g \int \text{d}^3 r \; \psi_\text{F}^{\dagger}(\mathbf{r})\psi_\text{B}^{\dagger}(\mathbf{r}) \psi_\text{B}(\mathbf{r}) \psi_\text{F}(\mathbf{r}),
\end{align} 
where $\psi_\text{B}(\mathbf{r})$ and $\psi_\text{F}(\mathbf{r})$ is the field operator for the bosons and fermions respectively. The coupling strength is $g=2\pi a/m_r$,  with 
$a$ the Bose-Fermi scattering length and $m_r=m m_\text{B}/(m+m_\text{B})$ the reduced mass. For a deep optical lattice, the 
fermion field operator can be approximated by $\psi_\text{F}(\mathbf{r}) =\sum_i \phi_0(\mathbf{r}-\mathbf{r}_i) a_i$ where $\phi_0(\mathbf{r}-\mathbf{r}_i)$ is the 
lowest Wannier function for site $i$ located at ${\mathbf r}_i$.  We approximate the Wannier function by a Gaussian 
$\phi_0(\mathbf{r})=\exp(-r_\perp^2/2\ell_\perp^2-z^2/2\ell_z^2)/\pi^{3/4}\ell_\perp\sqrt{\ell_z}$ where
  ${\mathbf r}_\perp=(x,y)$ is a vector in the lattice plane, and  $\ell_z$ and $\ell_\perp$ are the oscillator lengths of the potential wells of the 2D lattice perpendicular and parallel to the
plane. Using this in Eq.\ (\ref{Hint}), we obtain 
\begin{align}
H_\text{int}=\frac{g}{\mathcal V} \sum_{i,\mathbf{k},\mathbf{q}} e^{-\tfrac{1}{4}\ell_z^2 q_z^2} \,
e^{-\tfrac{1}{4}\ell_L^2 q_\perp^2} \,
e^{-i\mathbf{q}_\perp \cdot \mathbf{r}_i} \;
b_{\mathbf{k}+\mathbf{q}}^{\dagger} b_\mathbf{k} a_i^{\dagger} a_i
\label{Hint2}
\end{align}
for the Bose-Fermi interaction, where  ${\mathcal V}$ is the system volume and $b_{\mathbf{k}}^{\dagger}$ creates a boson with momentum ${\mathbf k}$. We have the 
usual Bogoliubov relation $b_{\mathbf{k}}=u_{\mathbf k}\gamma_{\mathbf k}-v_{\mathbf k}\gamma^\dagger_{-{\mathbf k}}$
with $u_{\mathbf k}^2=[1+(\epsilon_k^\text{B}+4\pi n_0a_\text{B})/E_{\mathbf k}]/2$ and $v_{\mathbf k}^2=[-1+(\epsilon_k^\text{B}+4\pi n_0a_\text{B})/E_{\mathbf k}]/2$.

\section{effective Hamiltonian for the fermions}
Since the bosons live in 3D whereas the fermions are confined to a 2D lattice, we expect the BEC to be essentially unaffected  by the fermions. On the other hand, the fermions interact with 
each other via the bosons and we will in this section derive an effective Hamiltonian describing this.  One fermion will either attract or repel the bosons thereby 
changing the local density of the BEC, which is felt by the second fermion. This results in the induced interaction 
\begin{align}
V_\text{ind}(i,j,i\omega_q)=g^2\int\!\frac{d^3q}{(2\pi)^3} &e^{i{\mathbf q}_\perp\cdot({\mathbf r}_i-{\mathbf r}_j)}e^{-(\ell_z^2 q_z^2-\ell_\perp^2 q_\perp^2)/2}\nonumber\\
&\times \chi_{\text{B}}(\mathbf{q},i\omega_q)
 \label{Vind}
\end{align}
between the fermions, where ${\mathbf q}_\perp=(q_x,q_y)$ is a  2D momentum, ${\mathbf q}=(q_x,q_y,q_z)$ is a 3D momentum, 
  and  $i\omega_q$ is a bosonic Matsubara frequency. 
The density-density correlation function of the BEC is  
\begin{align}
\chi_{\text{B}}(\mathbf{q},i\omega_q) = \frac{q^2}{m_B} \frac{n_0}{(i\omega_q)^2-E_\mathbf{q}^2}.
\end{align}
Equation (\ref{Vind}) includes an integration over the $z$ component $q_z$ of the boson momentum, 
since this  is not conserved by the Bose-Fermi interaction given by Eq.\ (\ref{Hint2}). 

For simplicity, we take the limit $\ell_\perp\rightarrow0$ and $\ell_z\rightarrow0$ in the following. 
We furthermore assume that the speed of sound $c_s=\sqrt{4\pi a_\text{B}n_0}/m_\text{B}$ in the BEC is much larger than $t a$, where $a$ is the lattice 
constant. This means that retardation effects are negligible so that the frequency can be set to zero in the induced interaction. 
When $c_s$ becomes comparable to $ta$, we expect retardation effects to significantly suppress pairing in analogy with what happens for the corresponding  system 
without a lattice~\cite{Wu2016}. Ignoring retardation effects by setting $i\omega_q=0$ in Eq.\ (\ref{Vind}) yields the usual Yukawa interaction~\cite{Viverit2000,Bijlsma2000} 
\begin{align}
V_\text{ind}(i,j) = -g^2\frac{n_0 m_B}{\pi} \frac{e^{-\sqrt{2}|\mathbf{r}_i-\mathbf{r}_j|/\xi_{\text{B}}}}{|\mathbf{r}_i-\mathbf{r}_j|},
\label{Vindr}
\end{align}
where $\xi_\text{B} = (8\pi a_{\text{B}} n_0)^{-1/2}$ is the BEC coherence length. It follows from Eq.\ (\ref{Vindr}) that the dimensionless parameter determining the 
strength of the induced interaction between the fermions in the lattice is 
\begin{align}
G= \frac{g^2  n_0 m_B}{ \pi at}.
\end{align}
Equation (\ref{Vindr}) illustrates another important fact: 
By varying the Bose density $n_0$ and/or the scattering lengths  $a_\text{B}$ and $a_\text{I}$, one can experimentally 
control both the \emph{strength} as well as the \emph{range} (determined by $\xi_\text{B}$) of the induced interaction between the fermions. 
%Below, we will show how this very appealing feature can be used to optimise the region in the phase diagram where the fermions form a topological superfluid.

Using the induced interaction, the effective Hamiltonian for the fermions is  
\begin{align}
H_\text{eff} = - t\sum_{\langle i,j \rangle} a_i^\dagger a_j-\mu\sum_ja_j^\dagger a_j +\sum_{i< j} V_\text{ind}(i,j) \, a_i^\dagger a_j^\dagger a_j a_i,
\label{Heff}
\end{align}
where $V_\text{ind}(i,j)$  is given by Eq.\ (\ref{Vindr}) and we have subtracted the chemical potential $\mu$ as usual.
Note that the system is  symmetric under the particle-hole transformation $\mathcal{P}a_i\mathcal{P}^{-1} = a^\dagger_i$, where the filling fraction 
transforms as $n\rightarrow 1-n$, the hopping matrix element as $t\rightarrow-t$, and  the chemical potential as $\mu\rightarrow-\mu+\sum_{j}V_\text{ind}(0,j)$.

\section{Zero temperature phase diagram}
Using the effective Hamiltonian given by Eq.\ (\ref{Heff}), we will  now calculate the  $T=0$ phase diagram of the fermions in the lattice. 
  We shall consider two possible instabilities of the system caused by the induced interaction: A superfluid  and a phase separation instability. To do this, we decouple the interaction  
  in the Hartree-Fock and BCS channels. Assuming pairing between $\mathbf{k}$ and $-\mathbf{k}$ states, the mean-field Hamiltonian becomes (apart from a constant)
%\begin{align}
%H_\text{MF} = \sum_\mathbf{k} \xi_\mathbf{k} a_\mathbf{k}^\dagger a_\mathbf{k} + \frac{1}{2}\left( \Delta_\mathbf{k}^\ast a_{-\mathbf{k}} a_\mathbf{k} + \Delta_\mathbf{k} a_{\mathbf{k}}^\dagger a_{-\mathbf{k}}^\dagger \right)
%\label{eq:grandpotential}
%\end{align}
\begin{align}
H_\text{MF}=\frac12\sum_\mathbf{k}
\begin{bmatrix}
a^\dagger_{\mathbf k} & a_{-\mathbf k}
\end{bmatrix}
\begin{bmatrix}
\xi_\mathbf{k} & \Delta_\mathbf{k}\\
 \Delta_\mathbf{k}^\ast & -\xi_\mathbf{k}
\end{bmatrix}
\begin{bmatrix}
 a_{\mathbf k} \\ a_{-\mathbf k}^\dagger
\end{bmatrix},
\label{eq:grandpotential}
\end{align}
where $\xi_\mathbf{k} = \epsilon_\mathbf{k}+\Sigma_\mathbf{k} -\mu$. Here,   
$\epsilon_\mathbf{k} = -2t \left( \cos k_x a +\cos k_y a \right)$ is the kinetic energy dispersion of the 2D lattice and 
\begin{align}
\Sigma_\mathbf{k} = \frac{1}{2\mathcal{V}} \sum_\mathbf{k'}&[V_\text{ind}(0)-V_\text{ind}(\mathbf{k}-\mathbf{k'})]\nonumber\\
& \times\left[ 1-\frac{\xi_\mathbf{k'}}{E_\mathbf{k'}} \tanh \left( \frac{E_\mathbf{k'}}{2T}  \right) \right]
\label{eq:sigmaselfcons}
\end{align}
is the Hartree-Fock self-energy. The Fourier transform  $V_\text{ind}(\mathbf{k})$ is given by 
\begin{align}
V_\text{ind}(\mathbf{k}) = \frac{1}{N} \sum_{i \neq 0} V_\text{ind}(0,i) e^{i \mathbf{k}\cdot \mathbf{r}_i},
\label{Vindk}
\end{align}
where $N$ is the number of lattice sites. 
The quasiparticle dispersion is $E_\mathbf{k} = (\xi_\mathbf{k}^2 + |\Delta_\mathbf{k}|^2)^{1/2}$ with the 
 gap parameter determined by 
\begin{align}
\Delta_\mathbf{k} = -\frac{1}{2\mathcal{V}} \sum_\mathbf{k'} V_\text{ind}(\mathbf{k}-\mathbf{k'}) \frac{\Delta_\mathbf{k'}}{E_\mathbf{k'}} \tanh \left( \frac{E_\mathbf{k'}}{2T}  \right). 
\label{eq:deltaselfcons}
\end{align}
As usual, we solve Eqs.\ (\ref{eq:sigmaselfcons}) and (\ref{eq:deltaselfcons}) self-consistently together with the number equation 
\begin{align}
n = \frac{1}{2} \left[ 1- \frac{1}{N} \sum_\mathbf{k} \frac{\xi_\mathbf{k}}{E_\mathbf{k}} \tanh \left( \frac{E_\mathbf{k}}{2T}  \right) \right],
\end{align}
which gives the filling fraction $0\le n\le 1$ of the lattice. Our numerical calculations are performed on a $N=121\times121$ lattice of $\mathbf{k}$-values in the first Brillouin zone.

\subsection{Topological $p_x+ip_y$ superfluid}
\begin{figure}[htb]
\centering
\includegraphics[width=\columnwidth]{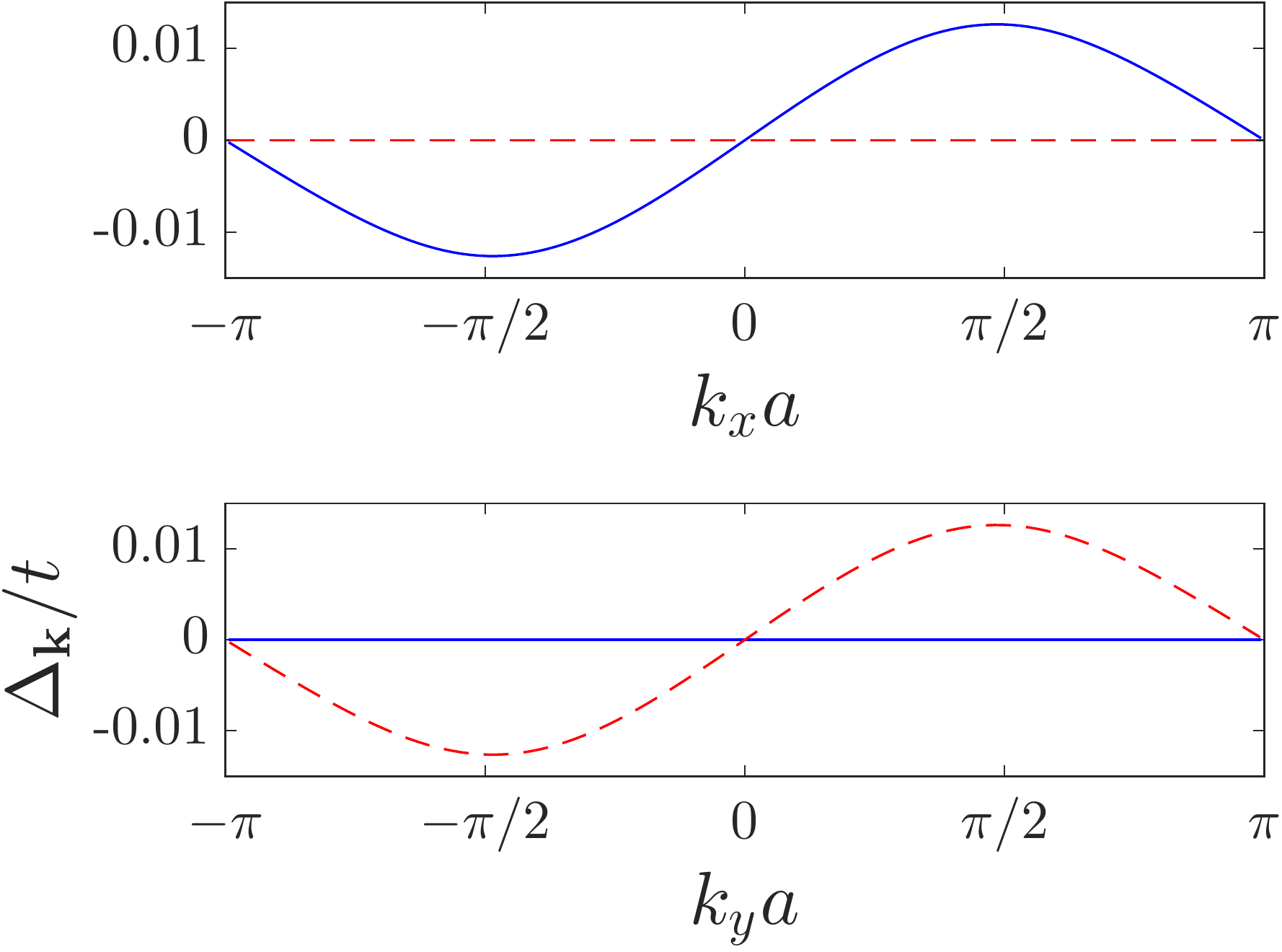}
\caption{The  gap parameter $\Delta_\mathbf{k}$ as a function of momentum $\mathbf{k}$ for the coupling strength $G=3$, BEC coherence length $\xi_B/a = 1$,  and filling fraction $n = 0.2$. 
The solid (blue) line shows the real part of the gap parameter, while the dashed (red) line shows the imaginary part. In the upper figure, we have taken $k_y=0$ 
and in the lower we have taken $k_x = 0$.} 
\label{fig:symmetry}
\end{figure}

At zero temperature, we find a solution characterised by a gap with $p_x+ip_y$ symmetry. Indeed,  the gap is  very  close to the pure $l=1$ angular momentum form 
$\Delta_{\mathbf k}\propto\sin k_xa+i\sin k_ya$ as illustrated in Fig.\ \ref{fig:symmetry}. 
%%%%%%%%%%%%%%%%%%%%%%%%%%%%%%%%%%%%%%%%%%%%%%%%%%%%%%%%%%%%%%
%%%%%%%%%%%%%%%%%%%%%%%%%%%%%%%%%%%%%%%%%%%%%%%%%%%%%%%%%%%%%%
This solution is the most stable energetically, as it fully gaps the Fermi surface, in contrast to for instance a solution with  $p_x$ symmetry~\cite{Anderson1961}. 
The $p_x+ip_y$ pairing  breaks time-reversal symmetry and  it is a class D topological superfluid with Majorana modes at its 
edges~\cite{Altland1997,Schnyder2008,Kitaev2009,Alicea2012}. The topological invariant is the Chern number  
\begin{align}
\nu=
\frac 1{4\pi}\int_\text{BZ}\!d^2 k \left[{\hat{\mathbf s}}(\mathbf{k})\cdot \left(\partial_{k_x}{\hat{\mathbf s}}(\mathbf{k})\times\partial_{k_y}{\hat{\mathbf s}}(\mathbf{k}) \right) \right],
\label{Chern2level}
\end{align}
where ${\hat{\mathbf s}}(\mathbf{k})={\mathbf S}(\mathbf{k})/|{\mathbf S(\mathbf{k}})|$. 
Here ${\mathbf S}(\mathbf{k})$ is defined as usual  by writing the BCS Hamiltonian as  
$\left[\begin{smallmatrix} \xi_\mathbf{k} & \Delta_\mathbf{k}\\
 \Delta_\mathbf{k}^\ast & -\xi_\mathbf{k}\end{smallmatrix} \right]={\mathbf S}(\mathbf{k})\cdot{\mathbf \sigma}$,
 with ${\mathbf \sigma}=( \sigma_x, \sigma_y, \sigma_z)$ the vector of  Dirac spin $1/2$ matrices.
 For our model, the Chern number is $\nu=-1$ for a filling fraction $0<n<1/2$ and $\nu=1$ for $1/2<n<1$. There is therefore a topological phase transition at half filling, $n=1/2$, where the spectral gap closes at the points $(k_x,k_y)=(0,\pm \pi/a)$ and $(k_x,k_y)=(\pm \pi/a,0)$ in the 
 Brillouin zone.

\subsection{Phase separation}
\begin{figure}[htb]
\centering
\includegraphics[width=\columnwidth]{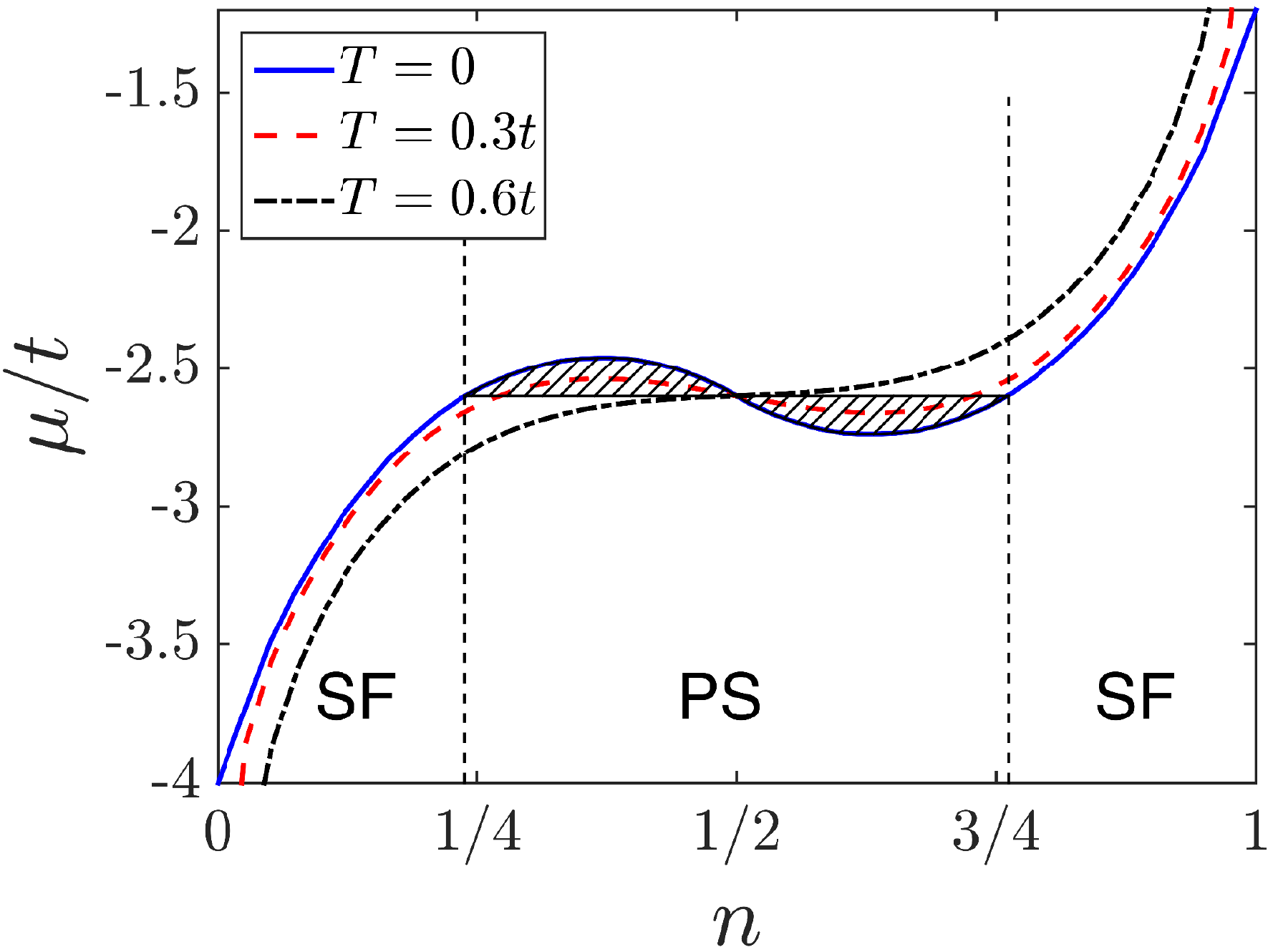}
\caption{Plot of the chemical potential $\mu$ as a function of filling fraction $n$ for $G=3$, $\xi_B/a=1$, and  different temperatures. The solid (blue) curve for $T=0$ has the Maxwell construction indicated, where the system is unstable towards phase separation for the filling fractions between the vertical dashed lines, and superfluid outside that region. The dashed (red) curve for $T=0.3t$ and the dot-dashed (black) curve for $T=0.6t$ show how the phase separation region shrinks and finally disappears with increasing  temperature.} 
\label{fig:chempot}
\end{figure}

The system becomes unstable towards phase separation
when the induced interaction between the fermions is too attractive, which originates from the  underlying particle-hole symmetry due to the lattice. The instability arises from the compressibility
$\kappa=n^{-2}\partial_\mu n$ being negative for certain filling fractions. As an example, we plot  in Fig. \ref{fig:chempot} the chemical potential  
 $\mu$ as a function of filling fraction $n$ for coupling strength $G=3$, range $\xi_B/a=1$, and  temperatures $T=0$, $T=0.3t$, and $T=0.6t$. For $T=0$ and $T=0.3t$, we see that $\mu$ 
 decreases with $n$ for a range of filling fractions, which corresponds to a negative compressibility, signalling that  the system is 
unstable towards phase separation.  The region of phase separation can  be determined 
by the Maxwell construction. Due to the particle-hole symmetry, this simplifies into the condition that a system with average filling fraction $n$ 
 phase separates into regions with filling fractions $n_1<n$ and $n_2=1-n_1>n$ determined by $\mu(n_1)=\mu(1/2)=\mu(n_2)$. The resulting 
 range of filling fractions where the system phase separates, is  shown in Fig. \ref{fig:chempot} for $T=0$.
 We also see from Fig. \ref{fig:chempot} that this range shrinks with increasing temperature. Indeed,  the system does not separate at all for $T=0.6t$. 
 The reason is that  the entropy of mixing stabilises the system against phase separation for a non-zero temperature. 
%%%%%%%%%%%%%%%%%%%%%%%%%%%%%%%%%%%%%%%%%%%%%%%%%%%%%%%%%%%%%%
%%%%%%%%%%%%%%%%%%%%%%%%%%%%%%%%%%%%%%%%%%%%%%%%%%%%%%%%%%%%%%

\subsection{Phase diagrams}
\begin{figure*}[htb]
\centering
\includegraphics[width=\textwidth]{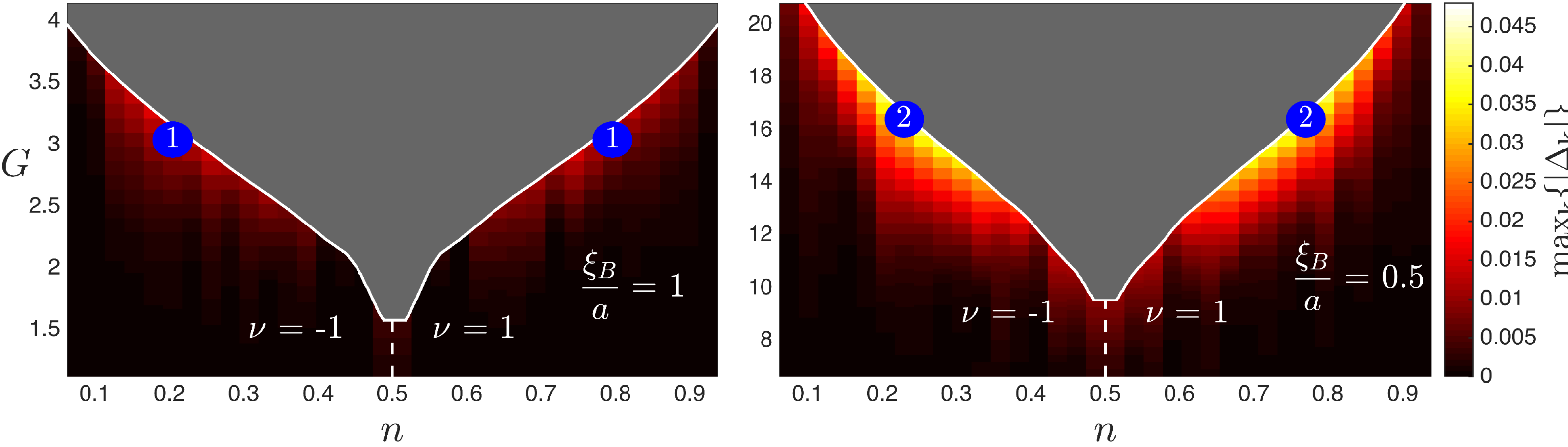}
\caption{Zero temperature phase diagram of the fermions as a function of filling fraction $n$ and coupling strength $G$ for $\xi_\text{B}/a=1$ (left) and $\xi_\text{B}/a=1/2$ (right).
 The color code indicates the maximal magnitude of the gap parameter in the Brillouin zone for a given set of 
$(n, G)$. The grey central regions indicate phase separation, and they are centered around half filling. %Below these regions we find no phase separation. 
 The numbered circles indicates the values $(n,G)$ where the pairing is maximal in the phase diagram. The kinks in the bottom of the phase separation regions are due to the finite resolution of the $G$-axis. An inspection of the phase separation condition shows that the region boundary must be smooth and have a vanishing derivative at the bottom. The vertical dashed lines 
 indicate a topological phase transition between a phase with Chern number $\nu=-1$ and a phase with $\nu=1$.}
\label{fig:heatmap}
\end{figure*}

We  now present  $T=0$ phase diagrams taking the pairing and phase separation instabilities into account. In Fig.\ \ref{fig:heatmap},
we plot phase diagrams as a function of the filling fraction $n$ and the Bose-Fermi coupling strength $G$
for two different values of the BEC coherence length: $\xi_\text{B}/a=1$ and $\xi_\text{B}/a=1/2$. The system is phase separated 
in the gray regions, whereas it is in the $p_x+ip_y$ superfluid state in the other regions with the color code indicating the maximum 
value of  $|\Delta_{\mathbf k}|$ in the Brillouin zone. The vertical line at half filling indicates a topological phase transition between a 
superfluid state with Chern number $\nu=-1$ and $\nu=1$. 
As expected, an increasing coupling strength $G$ increases the pairing. However, it also increases the 
range of densities where the system phase separates. Because of this competition, it is not simply a matter of increasing $G$ in order to increase the 
 pairing. If the attraction becomes too strong, the system simply phase separates into regions with filling fractions close to $n=0$ and $n=1$, which strongly 
suppresses pairing. Instead, one has to tune $G$ to an intermediate value to optimise the pairing. For a given $G$, one should choose an average  filling fraction $n$
in the phase separated region, i.e.\ $n_1\le n\le1-n_1$. The system will then phase separate into two superfluid regions with filling fractions $n_1$ and $1-n_1$, which have the same  
 pairing strength. An important conclusion from Fig.\ \ref{fig:heatmap} is that a smaller coherence length $\xi_\text{B}$ allows one 
to obtain a larger pairing strength by tuning $(n,G)$. Indeed, the maximum  value of the pairing one can achieve by tuning $(n,G)$ is 
$\max_{\textbf k}|\Delta_{\mathbf k}|=0.0142 t$ for $(\xi_\text{B}/a,G)=(1,3)$ and $n_1<n<1-n_1$ with $n_1=0.2$, compared to 
 $\max_{\textbf k}|\Delta_{\mathbf k}|=0.0382t$ for 
$(\xi_\text{B}/a,G)=(1/2,16.4)$ and $n_1<n<1-n_1$ with $n_1=0.23$.
The positions $(n_1,G)$ and $(1-n_1,G)$ of the maxima are indicated by the numbered circles \textcircled{1} for $\xi_\text{B}/a=1$ and \textcircled{2} for $\xi_\text{B}/a=1/2$ 
in Fig.\ \ref{fig:heatmap}. The reason that one can achieve a larger pairing for smaller  $\xi_\text{B}/a$, is that it determines the range of the induced interaction. 
Since phase separation is mainly driven by  long range interactions whereas pairing
 is mainly driven by the nearest neighbour interactions, a smaller range will suppress pairing less than it suppresses  phase separation. As a result of this
delicate competition, a small coherence length effectively favours pairing, since it allows a stronger coupling strength before the system phase separates. This shows that 
our proposed system is very useful for realising a topological superfluid 
in a lattice, since it allows the tuning of both  the strength and the range of the induced interaction.
%%%%%%%%%%%%%%%%%%%%%%%%%%%%%%%%%%%%%%%%%%%%%%%%%%%%%%%%%%%%%%
%%%%%%%%%%%%%%%%%%%%%%%%%%%%%%%%%%%%%%%%%%%%%%%%%%%%%%%%%%%%%%

\section{Critical temperature of superfluid phase}
Since the Fermi system is 2D, the superfluid phase melts via the BKT 
mechanism~\cite{Berezinskii1972,Kosterlitz1973,Kosterlitz1974,ChaikinBook}. In this section, we calculate the  
critical temperature $T_\text{BKT}$ of this transition  following the approach of Ref.~\cite{Gadsbolle2012}. 
The melting is determined by the phase stiffness of the order parameter, which  can be calculated from the 
 free energy cost associated with imposing a phase twist on the system. If the overall phase (in addition to the $p_x+ip_y$ phase winding) 
 of the order parameter changes by a small amount $\delta\theta$ between neighbouring sites along the $x$-direction, 
 the corresponding change in the free energy is 
 \begin{align}
F_\theta-F_0=N\frac {J_x}2 \delta\theta^2,
\label{StifnessDef}
\end{align}
 where $J_x$ is the phase stiffness along the $x$-direction. Imposing such a linear phase twist on the system 
 is equivalent to using periodic boundary conditions on a system described by the gauge transformed Hamiltonian~\cite{Lieb2002}
 \begin{align}
H_\text{eff}(\theta)=e^{-i\tfrac{\delta\theta}2\sum_lx_l/a}H_\text{eff}e^{-i\tfrac{\delta\theta}2\sum_lx_l/a}.
\label{GaugeTransform}
\end{align}
Here, $x_l$ is the $x$-coordinate of particle $l$ and we gauge transform each particle with half the angle $\delta\theta/2$, since the 
order parameter $\Delta_{\mathbf k}$
involves two particles. From Eq.\ (\ref{StifnessDef}), it is sufficient to calculate the energy shift due to the gauge transformation 
to second order in $\delta\theta$ to determine the superfluid stiffness. 
Expanding Eq.\ (\ref{GaugeTransform}) to second order in $\delta\theta$, and calculating the corresponding corrections to the energy yields 
after a lengthy but straightforward calculation~\cite{Gadsbolle2012}
  \begin{align}
J_x=\frac{t}{2N}\sum_{\mathbf k}\left[n_{\mathbf k}\cos k_xa-\frac {2t}Tf_\mathbf{k}(1-f_\mathbf{k})\sin^2k_xa \right]
\label{Stifness}
\end{align}
for the superfluid stiffness along the $x$-direction. Here, 
 $n_\textbf{k}=\langle a^\dagger_\textbf{k}a_\textbf{k}\rangle=|u_\textbf{k}|^2f_\textbf{k}+|v_\textbf{k}|^2(1-f_\textbf{k})$ is the number of fermions with 
momentum ${\mathbf k}$ and $f_\textbf{k}=(\exp\beta E_\textbf{k}+1)^{-1}$ is the Fermi function. An equivalent formula holds for the phase stiffness along the 
$y$-direction. We find that $J_x=J_y=J$ for $p_x+ip_y$ pairing. The critical temperature for the 
superfluid phase can now be determined by the BKT condition~\cite{Berezinskii1972,Kosterlitz1973,Kosterlitz1974,ChaikinBook}
\begin{align}
J(T_\text{BKT})=\frac2\pi T_\text{BKT}.
\label{BKTcond}
\end{align}

\begin{figure}[htb]
\centering
\includegraphics[width=\columnwidth]{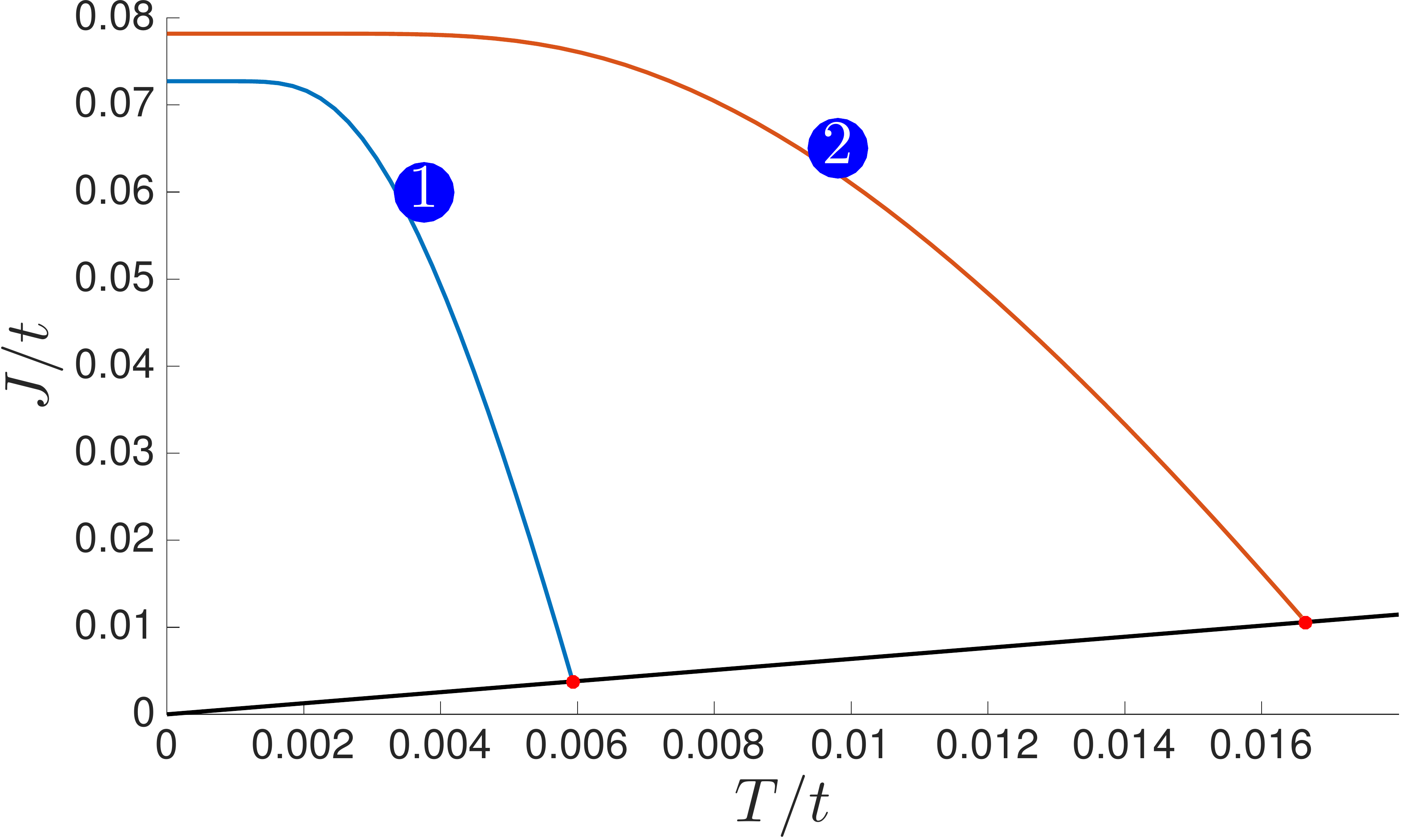}
\caption{The phase stiffness  given by Eq. (\ref{Stifness}) as a function of temperature for the two optimal parameter sets \textcircled{1} and \textcircled{2} shown in Fig. \ref{fig:heatmap}.
 The critical temperature is determined when $J(T)$ crosses the line $2T/\pi$, which is also plotted. }
\label{fig:BKT}
\end{figure}

In Fig.\ \ref{fig:BKT}, we plot the calculated phase stiffness $J(T)$ 
as a function of  temperature for the two different parameter sets \textcircled{1} and \textcircled{2} shown in Fig. \ref{fig:heatmap}. 
The critical temperature is from Eq. (\ref{BKTcond}) determined by the crossing of $J(T)$ with the line $2T/\pi$, which is also plotted in   Fig. \ref{fig:heatmap}. 
This condition yields the  BKT transition temperatures  $0.006t$ and $0.017t$ for 
$(\xi_\text{B}/a,n,G)=(1,0.2, 3)$ and $(\xi_\text{B}/a,n,G)=(1/2, 0.23,16.4)$ respectively. 
In these calculations, we use that the critical temperature $T_c$ of  the BEC typically is much larger than $T_\text{BKT}$, so that we can set $T=0$ when calculating the induced interaction 
mediated by the  BEC. It is  straightforward to include non-zero temperature effects on the BEC if necessary.  
%%%%%%%%%%%%%%%%%%%%%%%%%%%%%%%%%%%%%%%%%%%%%%%%%%%%%%%%%%%%%
%%%%%%%%%%%%%%%%%%%%%%%%%%%%%%%%%%%%%%%%%%%%%%%%%%%%%%%%%%%%%

\section{Discussion}
Despite the fact that  we have used the  flexibility of the Bose-Fermi system to optimise the interaction for topological pairing, the obtained 
 BKT critical temperatures are still fairly low.
 As an example, if one takes $^6$Li atoms in an optical lattice with a wavelength of $1000$nm and a lattice depth of $V_0 = 5 E_R$, the critical 
 temperature is $T_\text{BKT}\simeq2$nK using the parameters corresponding to \textcircled{2}  in Fig. \ref{fig:heatmap}. 
 The reason for this rather low temperature is that phase separation in the lattice prohibits the use of too strong  interaction. 
 One could of course obtain a higher absolute value of the critical temperature by tuning the lattice parameters or by decreasing 
 the coherence length, but here we have chosen to use commonly used experimental values as an example. Another 
 possibility is to use subwavelength lattices in order to increase the energy scales~\cite{Dubetsky2002,Nascimbene2015,Yi2008}.
 Finally, it is tempting to suppress  phase separation  by increasing the temperature, but as  we saw from 
the discussion in connection with Fig. \ref{fig:chempot},  the phase separation instability is unfortunately 
essentially unaffected by the low  temperatures $T\le T_{\rm BKT }$, since 
$T_{\rm BKT }\ll t$. We note however that  the critical temperatures one can obtain in the present system in general are higher than in other lattice proposals, since one can tune both the strength 
and the range of the interaction. For instance, we find a much lower critical temperature for the system of rotating dipoles in an optical lattice considered in Ref.~\cite{Liu2012},  
when the density is the same~\cite{Liu2012Comment}.

Since the Bose-Fermi mixture is a very tunable system compared to other proposals, the fact that we obtain rather low critical temperatures 
 indicates that it will be experimentally challenging to realise a topological superfluid using atoms in  an optical lattice. This should be compared with the similar Bose-Fermi system without a 
lattice, where high critical temperatures can be achieved~\cite{Wu2016}. 
The advantage however of using an optical lattice is  the available schemes for directly detecting the Majorana edge states. Current experiments with 
single site resolution~\cite{Sherson2010,Bakr2009} could specifically image the edge states using for instance their time-evolution in real space~\cite{Goldman2013,GoldmanEsslinger2016} or RF spectroscopy~\cite{Nascimbene2013}. Intriguingly, we note that since the two phase separated regions with filling fractions $n_1$ and $1-n_1$ have  Chern numbers 
$\nu=-1$ and $\nu=1$ respectively, there will be topologically robust edge states at the boundary between these two phases.

\section{Conclusions}
In this paper, we have analysed the phase diagram of identical fermions in a 2D square lattice immersed in a 3D BEC. The attractive induced interaction between the fermions mediated by 
the BEC was shown to give rise to topological $p_x+ip_y$ pairing as well as phase separation. We calculated the phase diagram at zero temperature as a function of the Bose-Fermi 
coupling strength and the filling fraction. The Bose-Fermi mixture was demonstrated to allow one 
to maximise topological superfluid pairing by tuning  the range of the 
interaction, so that it favours pairing between nearest neighbour fermions, while  long range interaction effects leading to phase separation are suppressed.  
We then calculated the BKT critical temperature for the superfluid phase, and we finally discussed the prospect for an experimental realisation of a topological superfluid in the present system.

\bibliographystyle{apsrev4-1}

\end{document}